\documentclass[aip,amsmath,apl,amssymb,reprint]{revtex4-2}
\usepackage{graphicx,color,amsmath,tabularx}
\usepackage{layouts}
\usepackage{etoolbox}
\usepackage[colorlinks=true,linkcolor=cyan,urlcolor=cyan,citecolor=cyan]{hyperref}
\usepackage[english]{babel}
\newcommand{\cm}{{cm$^{-1}$}}

\newcolumntype{C}[1]{>{\centering\arraybackslash}p{#1}} 

\begin{document}

\title{Substitutional sulfur and its vibrational fingerprints in Sb$_2$Se$_3$}

\author{F. Herklotz}
\email{frank.herklotz1@tu-dresden.de}
\affiliation{TU Dresden University of Technology, Institute of Applied Physics, 01062 Dresden, Germany}
\author{E. V. Lavrov}
\affiliation{TU Dresden University of Technology, Institute of Applied Physics, 01062 Dresden, Germany}
\author{A. Herklotz}
\affiliation{Institute of Physics, Martin-Luther-University Halle-Wittenberg, 06120 Halle, Germany}
\author{V.V. Melnikov}
\affiliation{Faculty of Physics, Tomsk State University, 634050 Tomsk, Russia}
\author{T. P. Shalvey}
\affiliation{Stephenson Institute for Renewable Energy, University of Liverpool, Liverpool L69 7ZF, UK}
\author{J. D. Major}
\affiliation{Stephenson Institute for Renewable Energy, University of Liverpool, Liverpool L69 7ZF, UK}
\author{K. Durose}
\affiliation{Stephenson Institute for Renewable Energy, University of Liverpool, Liverpool L69 7ZF, UK}

\date{\today}

\begin{abstract}
The configurational behavior of sulfur in antimony triselenide (Sb$_2$Se$_3$) is investigated by combining infrared absorption spectroscopy with density functional theory. Four sulfur-related local vibrational modes are identified at 249, 273, 283, and 312~cm$^{-1}$ in melt-grown single crystals prepared from Sb$_2$Se$_3$ granulate. Their assignment to sulfur is confirmed through controlled indiffusion experiments using Sb$_2$S$_3$ and elemental sulfur, as well as isotope-substitution studies with $^{34}$S, which produce the expected frequency shifts. Polarization-resolved measurements, together with theoretical calculations of local vibrational modes, demonstrate that the observed spectral features are fully consistent with substitutional sulfur on the three inequivalent selenium sites of Sb$_2$Se$_3$.
\end{abstract}

\maketitle

\section{Introduction}
In recent years, the antimony chalcogenides Sb$_2$Se$_3$, Sb$_2$S$_3$, and their solid solutions Sb$_2$(S$_{\rm x}$Se$_{\rm {1-x}}$)$_3$ have garnered significant attention as promising materials for a wide range of applications, including photovoltaics \cite{Chen_22_1}, photoelectrochemical devices and photocatalysis \cite{Yang_20_1}, photodetectors \cite{Zhai_10_1}, thermoelectrics \cite{Ko_16_1}, batteries \cite{Ou_17}, and phase-change memory devices \cite{Delaney_20_1}. Among these, Sb-based trichalcogenides are particularly notable as absorber materials for thin-film solar cells. Reported power conversion efficiencies have increased rapidly—from about 1\% in 2010 to 10.7\% in 2022 \cite{Zhao_22}—despite relatively limited research activity compared to more established materials like CdTe or CIGS \cite{Dale_23}.

\begin{figure}[t]
  \centering
    \includegraphics[width=6cm]{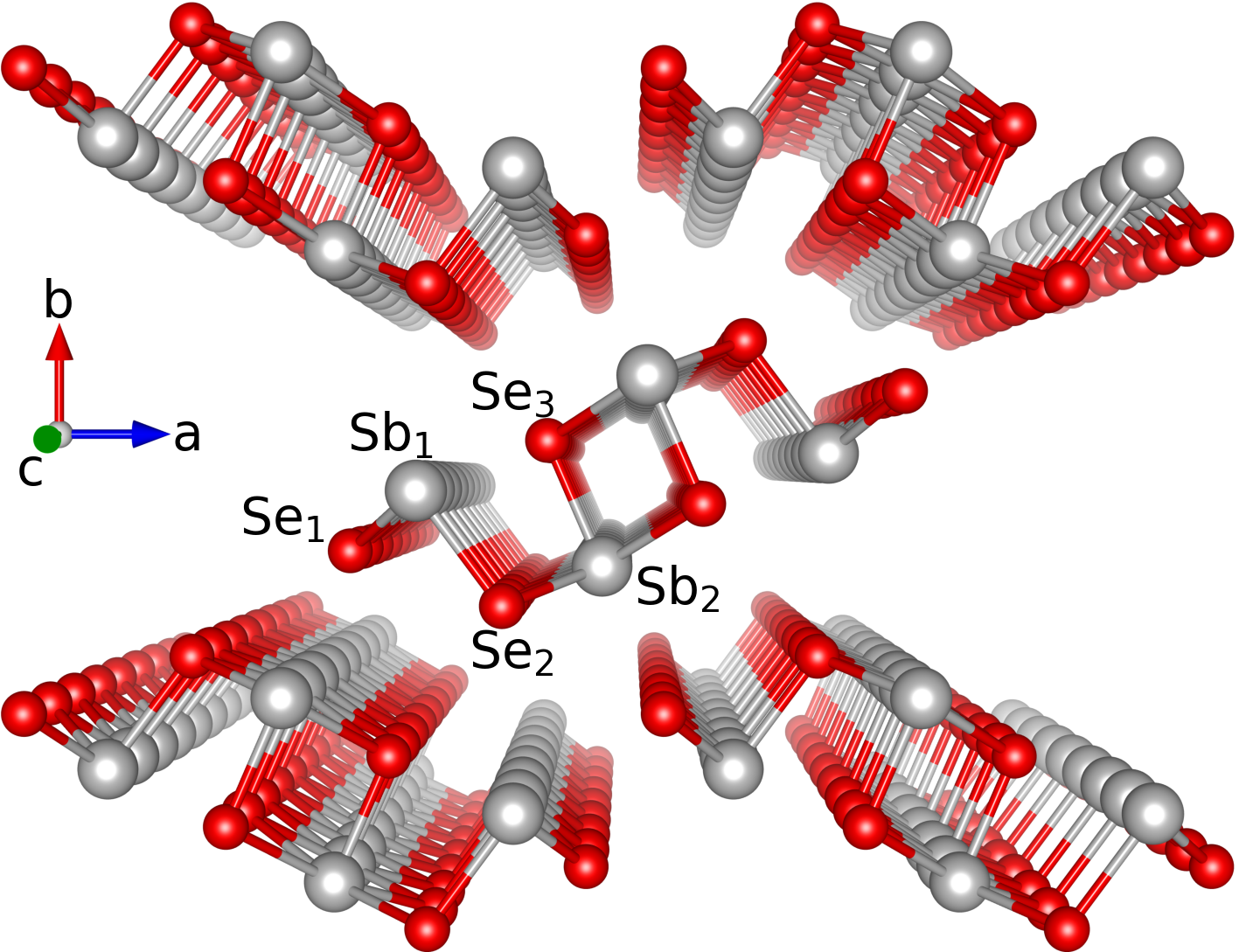}
    \caption{Structural model of Sb$_2$Se$_3$ viewed along the $c$ direction, indicating the three inequivalent Se sites. Red: Se, gray: Sb}  
\label{fig:DFT}
\end{figure}

A defining feature of Sb$_2$X$_3$ (X = S, Se) compounds is their quasi-one-dimensional orthorhombic \textit{Pbnm} crystal structure \cite{Tideswell_57,Doenges_50,Voutsas_85}, composed of covalently bonded [Sb$_4$X$_6$]$_n$ "nano-ribbons" oriented along the crystallographic $c$ axis that are held together by weak van der Waals interactions in the orthogonal directions (see Fig.~\ref{fig:DFT}), giving rise to highly anisotropic physical properties. This includes directionally dependent carrier transport, optical absorption, and dielectric response, distinguishing Sb$_2$Se$_3$ and Sb$_2$S$_3$ from traditional three-dimensional semiconductors and opening up new avenues for engineering directionally tailored device applications, provided the fundamental physical mechanisms are well understood.

Sulfur readily substitutes selenium in Sb$_2$Se$_3$ due to its comparable  chemical properties, smaller atomic radius (S: $\sim$1.84~\AA{} vs. Se: $\sim$1.98~\AA{}), and similar electronegativity. X-ray diffraction (XRD) studies consistently show a slight contraction in the lattice parameters ($a$, $b$, and $c$) with increasing sulfur content, consistent with the atomic size difference \cite{Yang_15_1,Khan_20_1,Pan_22_1}. Despite this contraction, the orthorhombic structure remains preserved, confirming that Sb$_2$(S$_{\rm x}$Se$_{\rm {1-x}}$)$_3$ forms a continuous solid solution over the full composition range ($0 \leq {\rm x} \leq 1$), without the appearance of secondary phases. Such alloying offers a tunable platform for tailoring electronic and optical properties, enabling band gap engineering and strain optimization for specific device applications. Moreover, sulfur incorporation has been shown to play a crucial role in the passivation of bulk defects in Sb$_2$Se$_3$ absorber layers, underscoring its significance for high-performance photovoltaic devices \cite{Prabhakar_22_1,Cai_24_1,Shu_25_1}.  

While the macroscopic properties of the Sb$_2$(S$_x$Se$_{1-x}$)$_3$ alloy system have been widely investigated, the local configurations and vibrational signatures of isolated sulfur atoms within the Sb$_2$Se$_3$ lattice remain poorly understood. Light dopants with masses smaller than those of the host atoms, such as sulfur substituting for selenium in Sb$_2$Se$_3$, often give rise to infrared-active local vibrational modes (LVMs)---vibrations that are decoupled from the extended phonon spectrum of the host lattice. Such LVMs provide sensitive probes of the microscopic nature of substitutional impurities in semiconductors, and their chemical identity can often be confirmed directly through the isotope mass dependence of the vibrational frequencies.

The unit cell of Sb$_2$Se$_3$ contains three crystallographically inequivalent selenium sites and two antimony sites (see Fig.~\ref{fig:DFT}), each embedded in a distinct local bonding environment. Among the selenium atoms, Se$_3$ is fivefold coordinated within the [Sb$_4$Se$_6$]$_n$ ribbon, forming one short covalent bond and four medium-range Se--Sb bonds \cite{Tideswell_57,Voutsas_85,Deringer_15_1}. By contrast, Se$_1$ and Se$_2$ are each threefold coordinated but differ strongly in geometry. Se$_1$ resides at the ribbon corner, with one covalent Sb--Se bond within the same unit and weaker van der Waals interactions to Sb atoms in neighboring ribbons. Se$_2$, on the other hand, is bonded to three Sb atoms within the same chain, comprising one short and two medium-range Sb--Se bonds. This diverse ensemble of local structural environments is expected to give rise to a rich set of distinct vibrational signatures, particularly when sulfur substitutes for selenium at these inequivalent sites.  

This work focuses on the characterization of specific microscopic configurations of sulfur incorporated into the Sb$_2$Se$_3$ lattice. By employing infrared absorption spectroscopy and density functional theory, we identify substitutional sulfur defects on the three inequivalent Se sites (S$_{\rm Se1}$, S$_{\rm Se2}$, S$_{\rm Se3}$) as the microscopic origin of sulfur-related local vibrational modes at 249, 273, 283, and 312~cm$^{-1}$. These findings provide a distinct spectroscopic fingerprint for isolated sulfur defects in Sb$_2$Se$_3$, offering a robust framework for characterizing and elucidating the role of sulfur in the solid solutions Sb$_2$(S,Se)$_3$. 

\section{Experimental and computational details}
\label{experiment}
The Sb$_2$Se$_3$ samples used in this study were two different types of single crystals grown by a vertical Bridgman melt-growth method, using identical thermal conditions, with the only distinction being the starting materials: i) high-purity elemental antimony (6N, Alfa Aesar) and selenium (5N, Alfa Aesar), and ii) commercially available Sb$_2$Se$_3$ granulate (5N, Alfa Aesar). These samples are hereafter referred to as ``elemental-grown'' and ``granulate-grown'', respectively. Additional details on the crystal growth and doping procedures are available in Refs.~\cite{Hobson_18,Hobson_20_3}.

The resulting ingots were typically 4~mm in diameter and 1--3~cm in length. Samples were prepared by cleaving and polishing thin slices with the surface normal oriented along either the crystallographic $b$ or $c$ axis. The final sample dimensions were approximately $3 \times 3$~mm$^2$ in area and $\sim$0.6~mm in thickness.

Infrared (IR) absorption spectra were recorded using a Bomem DA3.01 Fourier-transform infrared (FTIR) spectrometer, equipped with a globar source, a 3~$\mu$m Mylar beamsplitter, and a helium-cooled silicon bolometer detector. The spectral resolution was maintained between 1--2~cm$^{-1}$. Unless otherwise specified, measurements were performed on $b$-oriented samples, with the incident IR beam (k) aligned along the crystallographic $b$ axis. Polarization-dependent spectra were obtained using a wire-grid polarizer mounted on a KRS-5 substrate.

Samples were mounted in an Oxford OptistatCF helium exchange-gas cryostat equipped with polypropylene (outer) and high-density polyethylene (inner) windows. The sample temperature was regulated to within $\pm$1~K over the 12--300~K range and monitored using a DT-670 silicon diode sensor (Lake Shore), placed approximately 2~cm from the sample.

Post-growth sulfur incorporation into Sb$_2$Se$_3$ samples was performed via two diffusion methods. In the first, samples were sealed in evacuated quartz ampules with approximately 1~mg of sulfur powder (99.98\%, Sigma-Aldrich) and annealed at 180--400~$^\circ$C for 2~hours, followed by rapid quenching to room temperature ($\sim$2~min). This technique allows for the introduction of sulfur with natural abundances of the three dominant sulfur isotopes $^{32}$S (95\%), $^{33}$S (0.8\%), and  $^{34}$S (4.2\%) as well as isotopically-enriched sulfur. In the second method, a natural Sb$_2$S$_3$ crystal was co-sealed in the evacuated ampule, and the samples were annealed at 250--500~$^\circ$C under similar thermal and quenching conditions.

To estimate the local structure and vibrational frequencies of the S$_{\rm Se}$ defects in Sb$_2$Se$_3$, first principle calculations were performed within the framework of the density functional theory (DFT) using the projector augmented wave (PAW) method \cite{Bloechl_94_1} and the {PBEsol} exchange-correlation energy functional \cite{Perdew_08_1} as implemented in {Quantum ESPRESSO codes} \cite{Giannozzi_09_1,Giannozzi_17_1,pseudo_25_1}.
Under the  chosen level of theory the calculated lattice constants of bulk Sb$_2$Se$_3$, $a=$ 11.7848~\AA, $b=$ 3.9745~\AA, and $c=$ 11.3074~\AA, are in reasonable agreement with the corresponding experimental values of 11.7938, 3.9858, and 11.6478~\AA\ \cite{Voutsas_85}.

To construct the defect model for three possible configurations of S$_{\rm Se}$ a $a$\,$\times$\,3$b$\,$\times$\,$c$ orthorhombic supercell with periodic boundary conditions comprising 12 formula units was employed.
One selenium atom was substituted by sulfur one to create the substitutional defect under consideration.
All atomic positions were fully relaxed.

A series of high precision structure optimizations and subsequent calculations were performed using the $2\times 2\times 2$ grid of $k$-points.
The energy cut-offs were set to 120~Ry and 480~Ry for the plane wave basis set and the fine FFT grid, respectively.
The local vibrational modes associated with S$_{\rm Se}$ defects were calculated using the density-functional perturbation theory \cite{Baroni_01_1}.

\section{Results and discussion}

\subsection{Two-phonon transitions}
\label{2phonon}
The phonon spectrum of Sb$_2$Se$_3$ spans frequencies from approximately 30 to 215~cm$^{-1}$ \cite{VidalFuentes_19,Fleck_20_2}. Optical vibrational modes that predominantly involve displacements of Se atoms relative to the heavier Sb atoms typically occur above 100~cm$^{-1}$ \cite{Deringer_15_1}. By contrast, the lighter analogue Sb$_2$S$_3$ exhibits phonon frequencies extending up to $\sim$320~cm$^{-1}$ \cite{Petzelt_73_1,Ibanez_16_1}. Based on these considerations, we consider the frequency window of interest for local vibrational modes of substitutional sulfur defects in Sb$_2$Se$_3$ to lie between the upper limit of the intrinsic phonon spectrum ($\sim$215~cm$^{-1}$) and roughly 340~cm$^{-1}$. 
  
This spectral region, however, also contains a complex set of two-phonon absorption bands in Sb$_2$Se$_3$ \cite{Herklotz_25_2}. These features arise from anharmonic phonon--phonon interactions, in which two phonons are simultaneously created or annihilated during an optical transition. As a result, two-phonon absorption typically manifests as broad structures at frequencies corresponding to the sum or difference of fundamental phonon modes. Such bands can spectrally overlap with, and in some cases obscure, the sharper impurity-induced LVMs.
   
To reliably identify sulfur-related vibrational signatures, it is therefore essential to first establish the two-phonon background in high-purity reference crystals. For this purpose, we first analyze the infrared absorption spectra of nominally undoped ``elemental-grown'' Sb$_2$Se$_3$ single crystals. In Sec.~\ref{Smodes}, we then turn to the identification of sulfur-related LVMs in ``granulate-grown'' crystals, where sulfur impurities are unintentionally introduced by the granulate starting material.

\begin{figure}[t]
  \centering
    \includegraphics[width=0.48\textwidth]{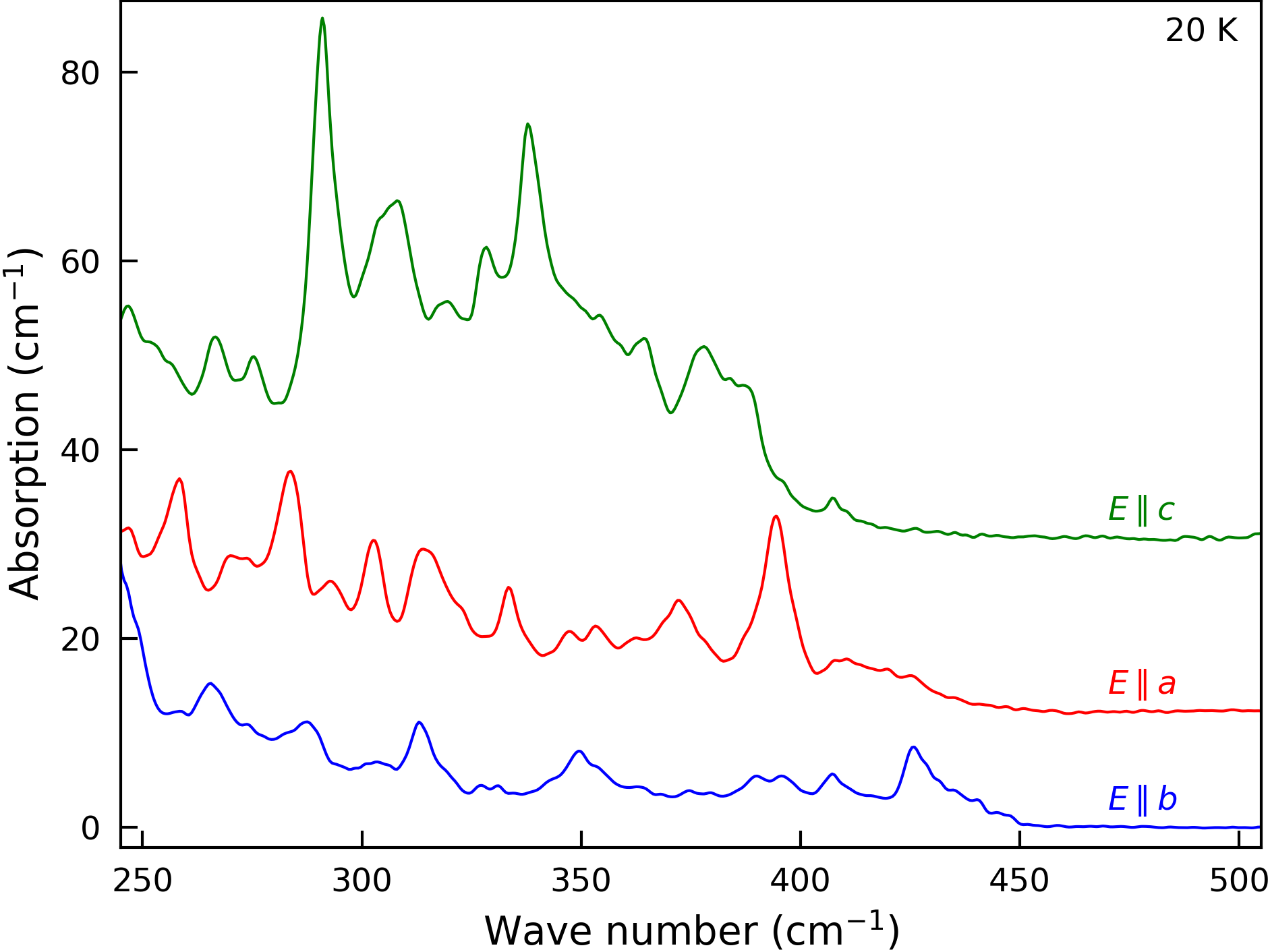}
    \caption{Polarized infrared absorption spectra in the two-phonon transition region, measured at 20~K for a nominally undoped ``elemental-grown'' Sb$_2$Se$_3$ single crystal. The spectra were recorded for different orientations of the electric field vector $E$ relative to the crystallographic axes. Baseline offsets were applied for clarity. The upper two spectra were obtained on a $b$-oriented sample, while the bottom spectrum was acquired on a $c$-oriented sample.}
\label{fig:2phonon}
\end{figure}

Figure~\ref{fig:2phonon} presents polarized IR absorption spectra of an undoped elemental-grown Sb$_2$Se$_3$ crystal recorded at 20~K with $E$ aligned along the three principal crystallographic axes. The spectra reveal a complex superposition of absorption bands, which we attribute to two-phonon processes. This rich spectral structure originates from the multiple inequivalent Se and Sb atomic sites, each associated with slightly different bonding environments \cite{Deringer_15_1}. The two-phonon bands extend up to $\sim$430~cm$^{-1}$ (see spectrum for $E \parallel b$), in line with approximately twice the highest optical phonon frequency of Sb$_2$Se$_3$ reported in the literature \cite{Petzelt_73_1,VidalFuentes_19,Fleck_20_2}.

A general trend is observed where the absorption intensity increases as the frequency decreases toward the range of fundamental phonon modes, which complicates IR measurements below $\sim$240~cm$^{-1}$ due to strong background absorption. Furthermore, the strength of the two-phonon absorption bands varies with polarization: relatively weak features are observed for $E \parallel b$ (blue), intermediate for $E \parallel a$ (red), and strong for $E \parallel c$ (green). This trend reflects the anisotropic bonding character in Sb$_2$Se$_3$, transitioning from van der Waals to increasingly covalent character along the $b$, $a$, and $c$ axes, respectively \cite{Deringer_15_1}.

\begin{figure}[t]
  \centering
    \includegraphics[width=0.48\textwidth]{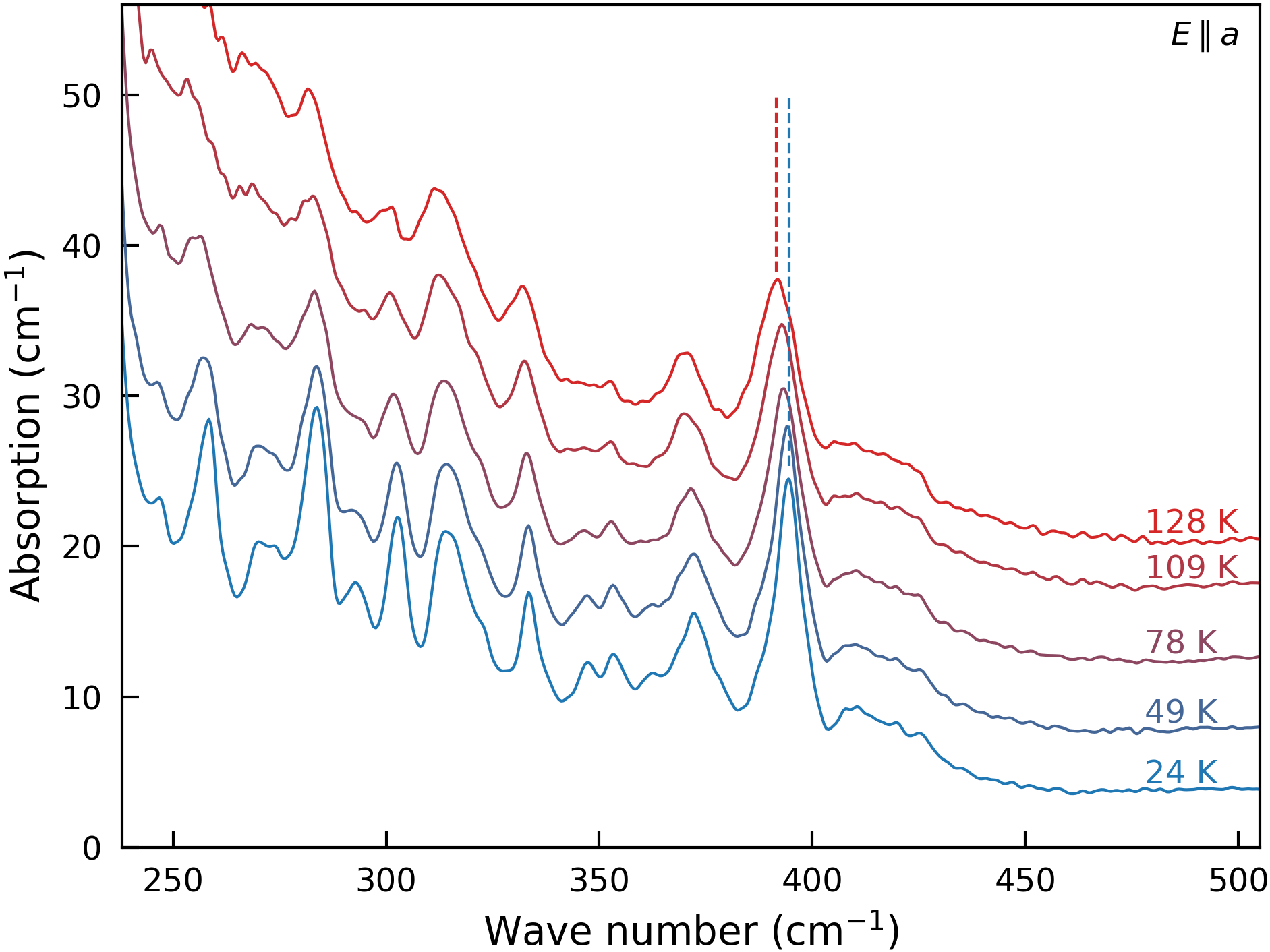}
    \caption{Temperature-dependent infrared absorption spectra in the two-phonon transition region, measured on a nominally undoped ``elemental-grown'' Sb$_2$Se$_3$ single crystal. The spectra were recorded with $k \parallel b$ and $E \parallel a$ and are offset for clarity.}
\label{fig:2phonon-Tdep}
\end{figure}

Figure~\ref{fig:2phonon-Tdep} presents temperature-dependent IR absorption spectra in the two-phonon region for light polarized along the $a$ axis. With increasing temperature, all observed bands broaden and experience a slight redshift. This down-shift in frequency is consistent with the known temperature dependence of the fundamental phonon modes \cite{VidalFuentes_19,Fleck_20_2}, reflecting thermal lattice expansion and a corresponding reduction in the force constants of the vibrating bonds.

\begin{table}[t]
 \caption{\label{tab:phonons} Peak positions at 20~K ({\cm}) of dominant two-phonon transitions of Sb$_2$Se$_3$ with their predominant polarization characteristics.}
 \begin{tabular}{C{0.135\textwidth}C{0.135\textwidth}C{0.135\textwidth}}\hline\hline
 $E \parallel$ a & $E \parallel$ b & $E \parallel$ c \\
  \hline
  \rule{0pt}{13pt} 258 &  266 & 267 \\
   285 & 288 &  275\\
   303 &  313 &  291 \\
   313 &  350 & 307 \\
   334 &  396 &  319 \\
   372 &  407 &  328 \\
   394 &  426 &  338 \\
  \hline\hline
 \end{tabular}
\end{table}

Table \ref{tab:phonons} summarizes the frequencies for the most dominant two-phonon transitions observed in $\text{Sb}_2\text{Se}_3$. It is noteworthy that the distinct polarization properties of these two-phonon absorption features allow for their use as an alternative non-destructive method for determining the orientation of bulk single crystals. This approach complements standard techniques like X-ray diffraction and Raman spectroscopy \cite{VidalFuentes_19,Fleck_20_2}. For instance, the prominent peaks at approximately 390~cm$^{-1}$ ($E \parallel a$), 430~cm$^{-1}$ ($E \parallel b$), and 290~cm$^{-1}$ ($E \parallel c$) can individually be used to assess the contributions of the respective crystallographic axes to the overall IR signal.
     
\subsection{Identification of local vibrational modes}
\label{Smodes}

\begin{figure}[t]
  \centering
    \includegraphics[width=0.48\textwidth]{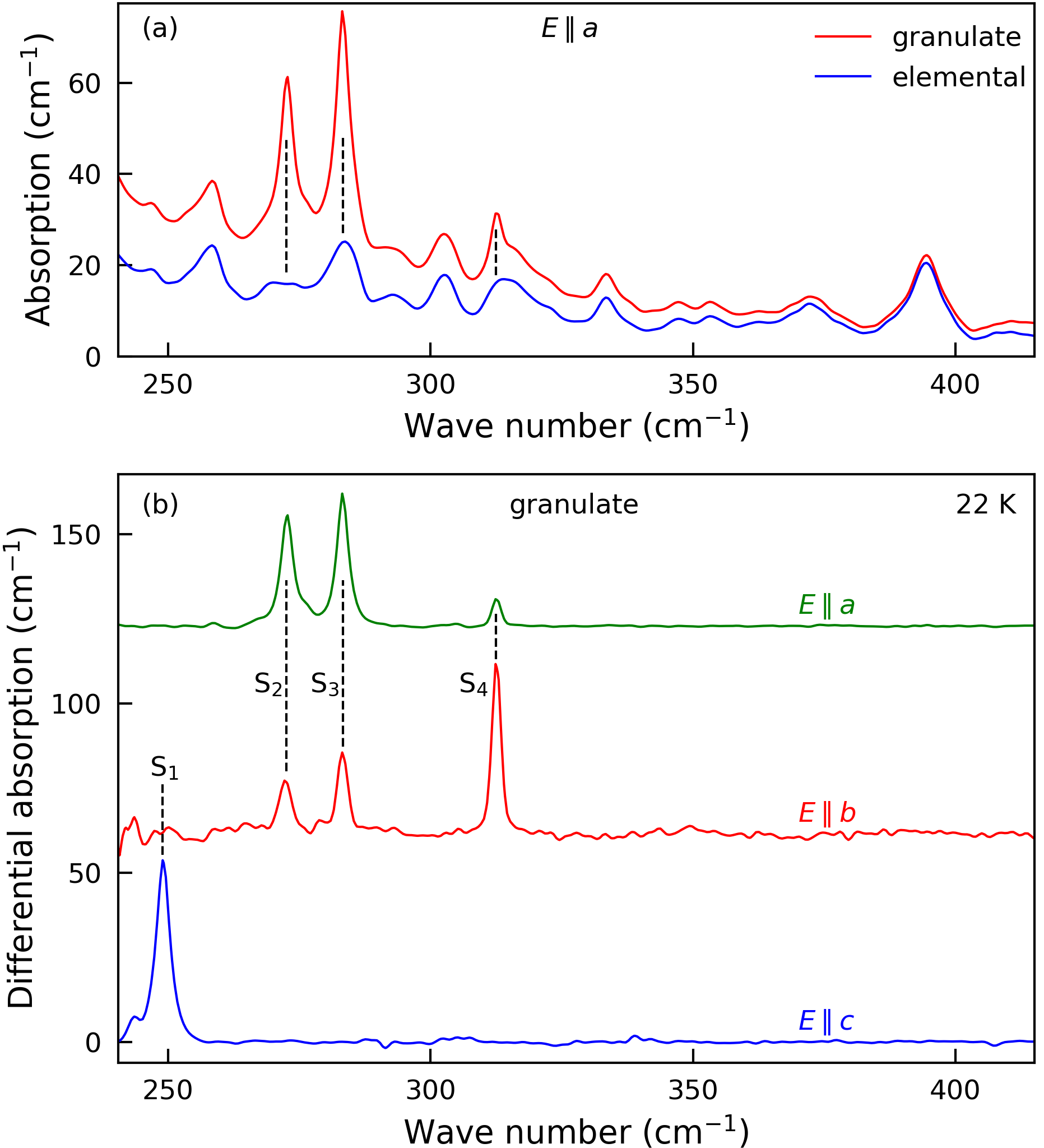}
    \caption{(a) Infrared absorption spectra of an ``elemental-grown'' (blue) and  ``granulate-grown'' (red) Sb$_2$Se$_3$ sample, recorded with $k \parallel b$ and $E \parallel a$. (b) Polarization-resolved differential absorption spectra of a granulate-grown sample, obtained by cross-subtracting spectra of an elemental-grown sample. Baseline offsets were applied for clarity. The upper and lower spectra were obtained on a $b$-oriented sample, while the mid spectrum was acquired on a $c$-oriented sample.}
\label{fig:peaks}
\end{figure}

Having thoroughly characterized the intrinsic two-phonon absorption of $\text{Sb}_2\text{Se}_3$, we now proceed to investigate impurity-related vibrational signatures. Figure \ref{fig:peaks} presents the results of IR absorption measurements on a granulate-grown $\text{Sb}_2\text{Se}_3$ crystal. Figure \ref{fig:peaks}(a) compares the absorption spectrum of this sample with that of an elemental-grown $\text{Sb}_2\text{Se}_3$ crystal, both recorded with light polarized parallel to the crystallographic $a$ axis. While both samples display the characteristic complex structure of two-phonon transitions detailed in the preceding section, the granulate-grown material additionally reveals a distinct substructure comprising three sharp lines at approximately 273, 283, and 312~cm$^{-1}$ (marked by dashed lines), which superimpose the intrinsic two-phonon spectrum.

To isolate these impurity-related features from the intrinsic background, we compute ``differential'' absorption spectra by subtracting the elemental-grown spectrum from that of the granulate-grown sample. Figure~\ref{fig:peaks}(b) shows these differential spectra for various polarization configurations. The impurity-induced lines are clearly visible as positive peaks in the difference spectra, confirming their localized character.

The polarization-resolved measurements reveal distinct anisotropies in these LVMs. The modes at 273 and 283~cm$^{-1}$ are essentially unpolarized within the $ab$ plane, whereas the 312~cm$^{-1}$ mode is strongly polarized along the crystallographic $b$ axis, indicating a more directional bonding environment. In addition, a fourth impurity-related mode appears at 249~cm$^{-1}$, observable only for light polarized along the $c$ axis. In the following, we denote these modes as S$_{1...4}$ in ascending order of frequency.  

\begin{figure}[t]
  \centering
    \includegraphics[width=0.48\textwidth]{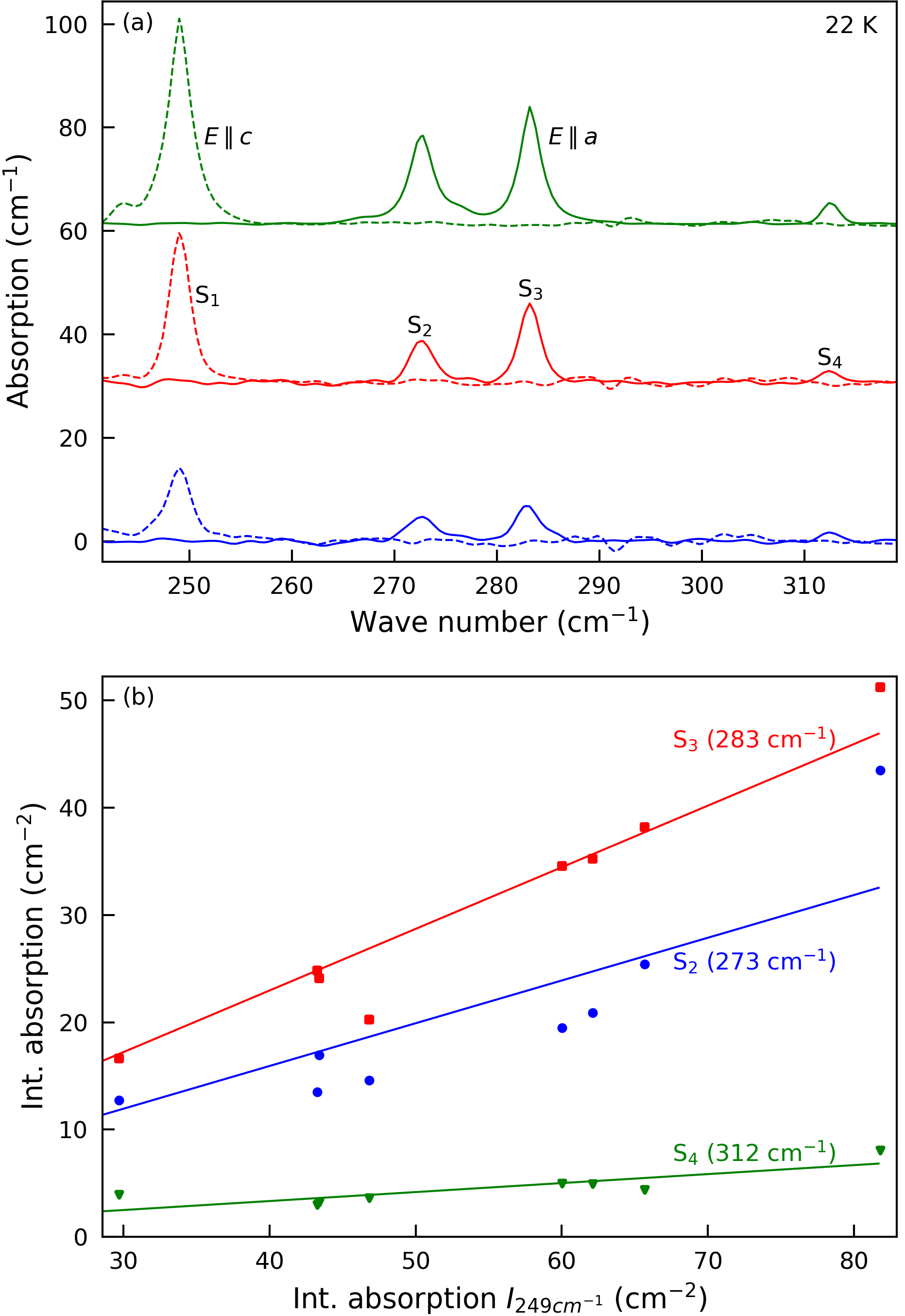}
    \caption{(a) Polarized differential absorption spectra of three granulate-grown Sb$_2$Se$_3$ samples. A nominally undoped elemental-grown Sb$_2$Se$_3$ crystal was used as the reference for subtraction. (b) Correlation between the integrated absorption coefficient of the 249~cm$^{-1}$ mode and those at 273, 283, and 312~cm$^{-1}$, obtained across eight granulate-grown Sb$_2$Se$_3$ samples.}
\label{fig:ratios}
\end{figure}

Figure~\ref{fig:ratios}(a) presents polarized differential absorption spectra of three granulate-grown Sb$_2$Se$_3$ samples. Panel (b) shows the correlation between the intensity of the S$_{2...4}$ modes and that of the S$_1$ line, measured across eight granulate-grown samples. Although the absolute intensities of the modes vary between samples, the relative intensities remain constant within experimental uncertainty. This strongly suggests that all four modes originate from a common type of defect or impurity. Indeed, the S$_{1...4}$ lines are observed consistently in all granulate-grown $\text{Sb}_2\text{Se}_3$ crystals, but are entirely absent in elemental-grown samples, demonstrating that the responsible impurities are likely introduced by the granulate starting material.  

In crystals, single impurity defects typically give rise to at most three vibrational modes. The observation of four distinct LVMs in the granulate-grown Sb$_2$Se$_3$ samples therefore points to a more complex scenario. The unit cell of Sb$_2$Se$_3$ contains three crystallographically distinct selenium sites, making it plausible that an impurity atom substitutes selenium at multiple inequivalent lattice positions. Such substitutional defects would generate a larger set of LVMs, some of which may fall within or near the strongly absorbing phonon region and thus remain experimentally inaccessible. For this scenario to be consistent with Fig.~\ref{fig:ratios}, the formation energies of the different substitutional configurations must be comparable, ensuring that the relative intensities of the S$_{1...4}$ lines remain nearly constant across samples.  

Support for this model is provided by the polarization properties of the S$_{1...4}$ modes. The site symmetry of substitutional defects at the three inequivalent Se sites is C$_{\rm s}$, implying a mirror plane perpendicular to the $c$ axis of the orthorhombic cell. Consequently, local vibrational modes of a single substitutional defect split into two A$'$ modes, polarized in the $ab$ plane, and one A$''$ mode, polarized along the $c$ axis. The experimental results follow this symmetry pattern: the S$_1$ mode at 249~cm$^{-1}$ corresponds to an A$''$ vibration, whereas the modes at 273, 283, and 312~cm$^{-1}$ correspond to A$'$ vibrations. Based on these considerations, we attribute the S$_{1...4}$ lines to a generic impurity X that substitutes for selenium at the three inequivalent lattice sites, i.e., X$_{\rm Se1}$, X$_{\rm Se2}$, and X$_{\rm Se3}$.  

\subsection{Sulfur introduction experiments}

The two most plausible candidates for the impurity X are chlorine and sulfur. Both elements are abundant, have atomic radii comparable to selenium, and can give rise to LVMs in the observed frequency range due to their atomic mass. Secondary ion mass spectrometry (SIMS) measurements indeed identified chlorine as a ubiquitous unintentional impurity in granulate-grown samples \cite{Hobson_20_2}. However, if chlorine were responsible, a characteristic doublet structure of the LVMs would be expected, reflecting the natural isotopic distribution of $^{35}$Cl (76\%) and $^{37}$Cl (24\%). No such splitting is observed. In the following section, we will demonstrate through controlled indiffusion experiments that the defects giving rise to the S$_{1...4}$ modes are instead associated with sulfur impurities.  

\begin{figure}[t]
  \centering
    \includegraphics[width=0.48\textwidth]{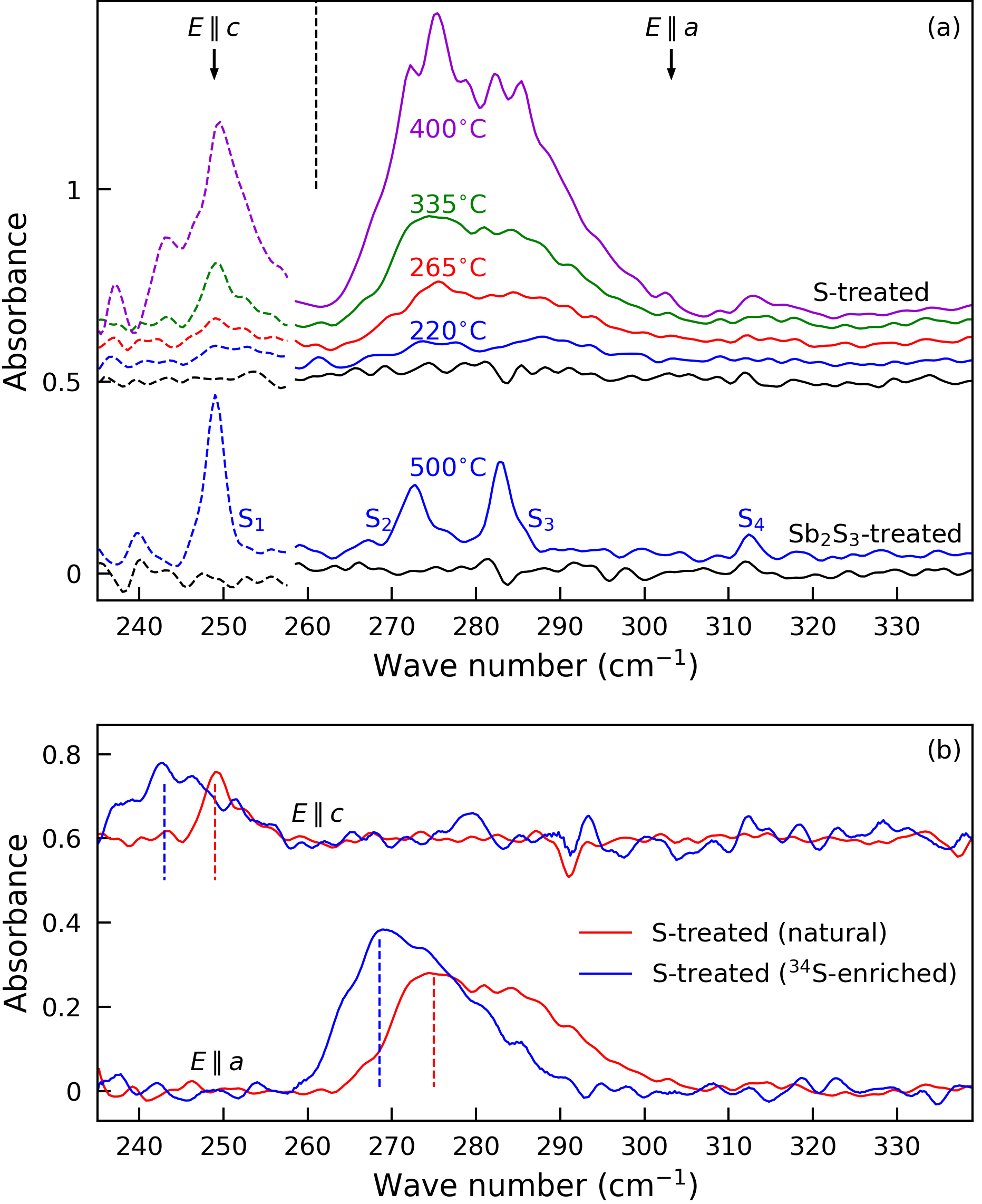}
  \caption{(a) Differential absorption spectra of two nominally undoped elemental-grown Sb$_2$Se$_3$ crystals subjected to post-growth treatments with sulfur and Sb$_2$S$_3$. Black curves correspond to the as-grown samples; colored curves show spectra after treatment at the indicated temperatures. (b) Polarized differential absorption spectra of Sb$_2$Se$_3$ samples treated with either sulfur of natural isotopic composition (red) or 95~at.\% $^{34}$S-enriched sulfur (blue).}
\label{fig:treatments}
\end{figure}

Figure \ref{fig:treatments} (a) displays differential absorption spectra acquired on two elemental-grown Sb$_2$Se$_3$ crystals subjected to a series of post-growth treatments with sulfur (top) and Sb$_2$S$_3$ (bottom) in sealed, evacuated quartz ampules (see Sect.~\ref{experiment} for details). The black curves correspond to the as-grown samples, while the colored curves show spectra after treatments at the indicated temperatures.  To enhance visibility, data below $\sim$260~cm$^{-1}$ are plotted for light polarized along the $c$ axis (dashed curves) to highlight the 249~cm$^{-1}$ line, whereas spectral data above 260~cm$^{-1}$ are shown for light polarized along the $a$ axis (solid curves) to access the 273, 283, and 312~cm$^{-1}$ modes. Data not shown do not display any additional spectral features.  

The as-grown crystals exhibit essentially flat baselines, confirming the absence of the defect modes of interest prior to treatment. After Sb$_2$S$_3$ treatment, four sharp features appear whose frequencies coincide with the S$_{1...4}$ modes observed in granulate-grown crystals. Their polarization behavior is also consistent: the 249~cm$^{-1}$ line is polarized parallel to the $c$ axis, while the 273, 283, and 312~cm$^{-1}$ modes are polarized perpendicular to $c$. These results strongly indicate that the S$_{1...4}$ modes in nominally undoped granulate-grown and Sb$_2$S$_3$-treated elemental-grown crystals share a common origin.

Treatment of Sb$_2$Se$_3$ crystals with elemental sulfur also produces additional features whose intensities increase with diffusion temperature. Compared to the S$_{1\ldots4}$ modes, these bands are significantly broader and blueshifted by about 1--3~cm$^{-1}$. We tentatively attribute this behavior to a higher concentration of near-surface sulfur defects induced by elemental sulfur treatment, which may generate enhanced local strain. Nevertheless, the polarization properties of these broader features remain unchanged: a single low-frequency mode polarized parallel to $c$, and higher-frequency modes polarized perpendicular to $c$.
      
Taken together, these results provide strong evidence for the sulfur-related nature of the S$_{1...4}$ modes. As a counter-check, annealing elemental-grown Sb$_2$Se$_3$ in an argon atmosphere produced none of the S$_{1...4}$ features.  

Additional confirmation comes from isotopic substitution. Figure~\ref{fig:treatments}(b) compares differential spectra of samples treated with natural sulfur (dominated by $^{32}$S) and with isotopically enriched $\approx$95~at.\% $^{34}$S, measured for light polarized both parallel and perpendicular to the $c$ axis. The $^{34}$S-treated spectra are consistently downshifted by 5--7~cm$^{-1}$ relative to the natural-sulfur case. The observed frequency ratios of the peak maxima, 243/249 = 0.976 and 268.5/275 = 0.976, closely match the value expected from the isotopic mass effect of $^{34}$S versus $^{32}$S, $\sqrt{32/34} \approx 0.97$. This provides direct evidence that the S$_{1\ldots4}$ modes originate from sulfur-related local vibrational modes.

\subsection{Temperature-dependence of the LVMs}

\begin{figure}[t]
  \centering
    \includegraphics[width=0.48\textwidth]{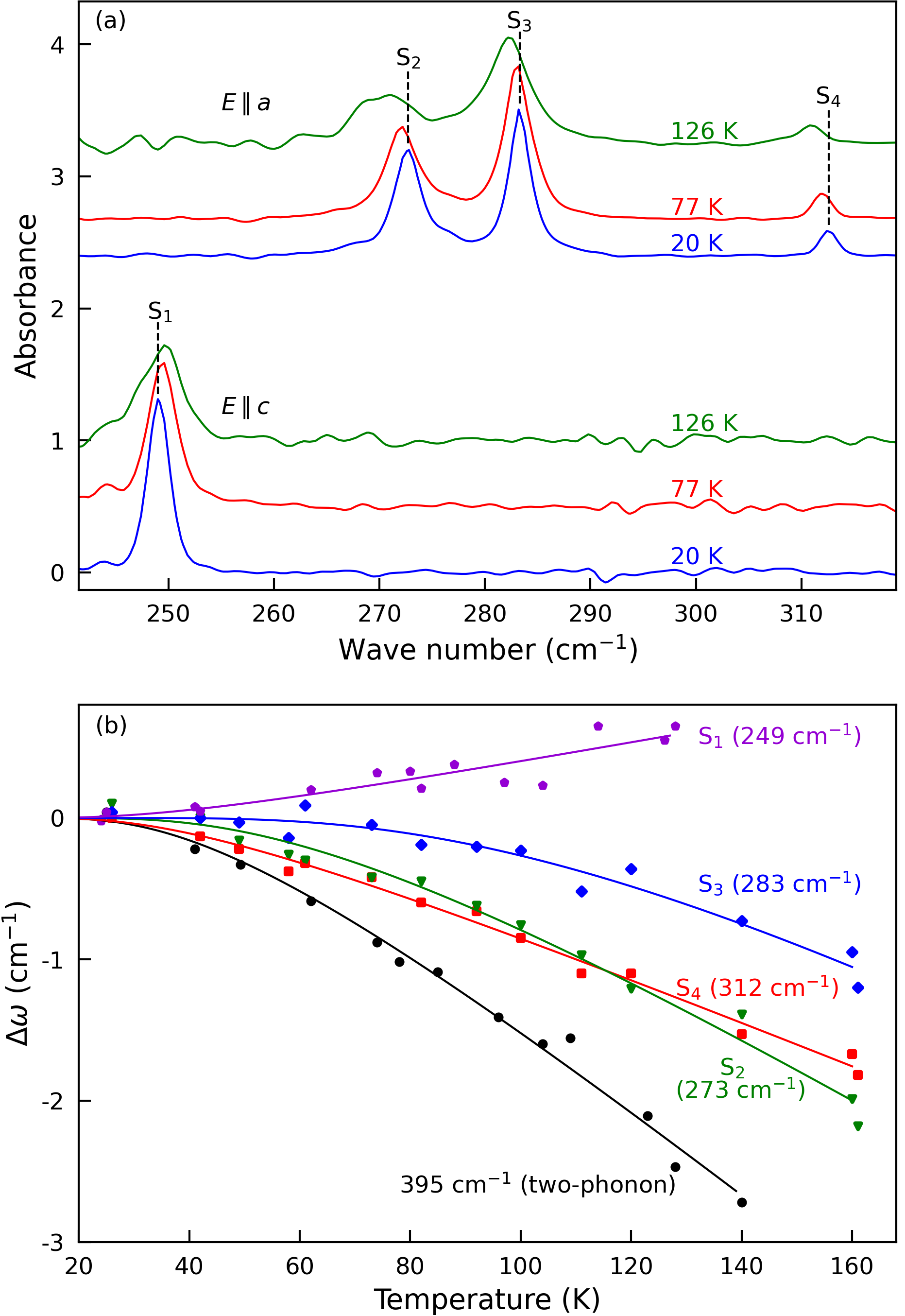}
    \caption{Temperature dependence of the absorption lines under study. (a) Polarized differential absorption spectra of a granulate-grown Sb$_2$Se$_3$ crystal, measured with $k \parallel b$. A nominally undoped elemental-grown Sb$_2$Se$_3$ single crystal served as the reference. (b) Peak positions of the S$_{1\ldots4}$ modes and of the two-phonon transition at 395~{\cm} as a function of temperature.}
\label{fig:Tshifts}
\end{figure}

Figure~\ref{fig:Tshifts} summarizes the temperature dependence of the sulfur-related vibrational modes in a nominally undoped, granulate-grown Sb$_2$Se$_3$ crystal. The upper panel shows representative polarized differential absorption spectra recorded at three distinct temperatures, while the lower panel tracks the peak positions of the S$_{1\ldots4}$ modes as a function of temperature. For comparison, the frequency shift of the dominant $b$-oriented two-phonon transition at $\sim$395~{\cm} (black symbols) is also included as a benchmark for intrinsic phonon-related behavior.  

To quantitatively describe the temperature-induced frequency shifts shown in the lower panel, we employ the model proposed by Persson and Ryberg \cite{Persson_85_1}, which accounts for anharmonic coupling of a harmonic vibration to an exchange mode of energy $E_0$. Within this framework, the shift $\Delta \omega$ of the vibrational frequency is given by  

\begin{equation}
\Delta \omega(T) = \frac{\delta \omega}{\exp(E_0/kT)-1},
\end{equation}

where $\delta \omega$ is the coupling constant. The solid lines in Fig.~\ref{fig:Tshifts} represent fits to this expression and reproduce the experimental data reasonably well, capturing both the magnitude and the trend of the temperature-induced shifts. The energies of the exchange mode $E_0$ obtained are in the range 74$-$220~{\cm}, indicating that the modes are coupled to lattice vibrations.

The analysis reveals that the temperature-induced shifts of the S$_{1\ldots4}$ modes are generally smaller in magnitude than those of the most intrinsic two-phonon transitions, consistent with the more localized character of LVMs, which makes them less sensitive to global lattice expansion. More importantly, however, the thermal behavior differs significantly among the individual sulfur-related modes, both in the fitted coupling constant $\delta \omega$ and in the activation energy $E_0$. The qualitative temperature dependence also varies: the higher-frequency modes S$_{2\ldots4}$, as well as the two-phonon transition, exhibit the expected redshift with increasing temperature, reflecting contributions from lattice expansion, softening of restoring forces, and enhanced phonon interactions. By contrast, the S$_1$ mode shows a distinct blueshift with increasing temperature.
 
We tentatively attribute this contrasting behavior to the polarization characteristics of the individual defect-related modes. Specifically, the S$_1$ mode is polarized along the crystallographic $c$ axis, corresponding to an out-of-plane vibration with $A''$ character, whereas the S$_{2\ldots4}$ lines are associated with $A'$-type vibrations representing in-plane motions within the $ab$ plane. Since Sb$_2$Se$_3$ possesses a highly anisotropic crystal structure, characterized by strong covalent bonds along the $c$ axis and weak van der Waals interactions perpendicular to it, a pronounced directional dependence of the phonon coupling for these defect modes is expected.  

Interestingly, similar anisotropic temperature responses of defect- or phonon-related modes have been reported in other layered chalcogenides, such as Sb$_2$S$_3$ and Bi$_2$Se$_3$, where out-of-plane and in-plane vibrations exhibit markedly different temperature coefficients \cite{Gan_15_1,Zhao_11_1}. This broader comparison underscores that the unusual blueshift of the S$_1$ mode in Sb$_2$Se$_3$ may reflect a more general phenomenon in quasi-one-dimensional or layered materials, where strongly anisotropic bonding environments lead to unconventional thermal behavior of localized vibrational modes.  

\subsection{Calculations}

The previous results provide strong evidence for the involvement of sulfur in the defects responsible for the S$_{1...4}$ lines at 249, 273, 283, and 312~{\cm}. Here, we present first-principles calculations of S$_{\rm Se}$ defects in Sb$_2$Se$_3$.
 
Our calculations reveal that S$_{\rm Se}$ defects possess low formation energies under Se-poor conditions, consistent with an earlier report on S-induced point defects \cite{Guo_19_3}. Moreover, the three inequivalent Se sites (S$_{\rm Se1}$, S$_{\rm Se2}$, S$_{\rm Se3}$, see Fig.~\ref{fig:DFT}) differ only marginally in energy: the S$_{\rm Se2}$ configuration is the most stable, while S$_{\rm Se3}$ and S$_{\rm Se1}$ are only 6.6 and 10.1~meV higher, respectively. These small energy differences suggest that sulfur can substitute for selenium nearly at random within the Sb$_2$Se$_3$ lattice, consistent with the well-established observation of continuous solid solution of Sb$_2$(S$_{\rm x}$Se$_{\rm {1-x}}$)$_3$.  

The calculated vibrational frequencies associated with the three inequivalent substitutional S$_{\rm Se}$ sites are summarized in Table~\ref{tab:vib}, together with the experimental peaks, their symmetry assignments, and the frequency differences between experiment and theory. Notably, applying a uniform upward shift of $\sim$23~{\cm} to the computed frequencies yields excellent agreement with the experimental data.  

It is well established that theoretical analyses of local vibrational modes are subject to systematic errors in the absolute frequency values. A similar systematic discrepancy to that observed here has recently been reported for phonon energies in Sb$_2$Se$_3$ and other chalcogenides using the PBEsol functional \cite{Fleck_20_2,Whittles_19_1}. Such deviations most likely originate from differences between computational and experimental lattice parameters, which arise from the athermal approximation and from limitations of density functional theory in accurately describing interatomic forces. For materials such as Sb$_2$Se$_3$, where strong covalent bonds coexist with weak van der Waals interactions along different crystallographic directions, these challenges are particularly pronounced.  

For reference, we also calculated phonon mode frequencies of Sb$_2$Se$_3$ and its isostructural counterpart Sb$_2$S$_3$. In Sb$_2$Se$_3$, excellent agreement with all experimental observed optical phonon modes above 150~{\cm} was obtained after applying a systematic upward shift of 11~{\cm}, consistent with earlier reports \cite{Fleck_20_2,VidalFuentes_19}. For Sb$_2$S$_3$, an even larger scaling factor of 18~{\cm} results in good agreement with the experimental high-frequency phonon modes \cite{Ibanez_16_1}.  

In light of these findings, we conclude that the uniform shift of $\sim$23~{\cm} observed for the S$_{\rm Se}$ defects in Sb$_2$Se$_3$ is well within reasonable expectations. The excellent agreement between experiment and theory after applying this correction strongly supports the assignment of the S$_{1...4}$ modes to substitutional S$_{\rm Se}$ defects.  

\begin{table}[t]
 \caption{\label{tab:vib} Local vibrational modes of the S$_{\rm Se}$ defects in Sb$_2$Se$_3$  (\cm). The Se site geometries are as indicated in Fig. \ref{fig:DFT}. Only experimentally accessible modes above the phonon background are shown.}
 \begin{tabular}{C{0.07\textwidth}C{0.07\textwidth}C{0.07\textwidth}C{0.07\textwidth}C{0.07\textwidth}}\hline\hline
 Defect & \multicolumn{2}{c}{Theory} & Experiment & $\Delta \omega$ \\
  \hline
  \rule{0pt}{13pt}S$_{\rm Se1}$ & $A'$  & 155 &  &  \\
   & $A''$  & 225 & 249 & 24 \\
   & $A'$  & 249 & 273 & 24 \\
   \hline
  S$_{\rm Se2}$ & $A''$  & 201 &  & \\
   & $A'$  & 222 &  & \\
   & $A'$  & 258 & 283 & 25\\
   \hline
  S$_{\rm Se3}$ & $A''$  & 175 &  &  \\
   & $A'$  & 175 &  &  \\
   & $A'$  & 292 & 312 & 20 \\
  \hline\hline
 \end{tabular}
\end{table}
 
\section{Conclusions}
In this work, we investigate the microscopic configurations of sulfur incorporated into the Sb$_2$Se$_3$ lattice by combining infrared absorption spectroscopy with density functional theory. We identify substitutional S$_{\rm Se}$ defects as the origin of four distinct sulfur-related local vibrational modes at 249, 273, 283, and 312~cm$^{-1}$. These results establish an unambiguous spectroscopic fingerprint for isolated sulfur defects in Sb$_2$Se$_3$, providing a framework for understanding the role of sulfur in the solid solutions Sb$_2$(S,Se)$_3$.

\section*{Acknowledgments }
This work was funded by the Deutsche Forschungsgemeinschaft (Grant No. LA 1397/21) as well as the Engineering and Physical Sciences Research Council grants EP/W03445X/1 and EP/M024768/1. VVM acknowledges the support from the Ministry of Education and Science of the Russian Federation.

%

\end{document}